\documentclass[conference]{IEEEtran}
\IEEEoverridecommandlockouts

\usepackage[noadjust]{cite}
\usepackage{amsmath,amssymb,amsfonts}
\usepackage{algorithm}
\usepackage{algorithmic}
\usepackage{graphicx}
\usepackage{textcomp}
\usepackage{xcolor}
\usepackage{amsthm}
\usepackage{xspace}
\usepackage{nicefrac}
\usepackage[bookmarks=false]{hyperref}
\usepackage{bbm}
\usepackage{caption}
\usepackage{subcaption}
\usepackage{flushend}
\usepackage[absolute,overlay]{textpos}

\newcommand{\Anon}{{$z$-anon}\xspace}
\newcommand{\kanon}{{$k$-anon}\xspace}

\newcommand{\comment}[1]{\textit{//#1}}

\begin{document}

\title{$z$-anonymity:\\ Zero-Delay Anonymization for Data Streams
\thanks{The research leading to these results has been funded by the European Union's Horizon 2020 research and innovation program under grant agreement No. 871370 (PIMCity) and the SmartData@PoliTO center for Big Data technologies.}
}
\author{Nikhil Jha, Thomas Favale, Luca Vassio, Martino Trevisan, Marco Mellia \\
\small{Politecnico di Torino}\\
\texttt{first.last@polito.it}
}

\maketitle

\TPshowboxestrue
\TPMargin{0.3cm}
\begin{textblock*}{15.5cm}(3cm,0.4cm)
\footnotesize
\bf
\definecolor{myRed}{rgb}{0.55,0,0}
\color{myRed}
\noindent
Please cite this article as: Nikhil Jha, Thomas Favale, Luca Vassio, Martino Trevisan, Marco Mellia. z-anonymity: Zero-Delay Anonymization for Data Streams. 2020 IEEE International Conference on Big Data (Big Data). DOI: \url{https://doi.org/10.1109/BigData50022.2020.9378422}
\end{textblock*}

\begin{abstract}
With the advent of big data and the birth of the data markets that sell personal information, individuals' privacy is of utmost importance. The classical response is anonymization, i.e., sanitizing the information that can directly or indirectly allow users' re-identification. 
The most popular solution in the literature is the $k$-anonymity. However, it is hard to achieve $k$-anonymity on a continuous stream of data, as well as when the number of dimensions becomes high. 

In this paper, we propose a novel anonymization property called $z$-anonymity. Differently from $k$-anonymity, it can be achieved with zero-delay on data streams and it is well suited for high dimensional data. The idea at the base of $z$-anonymity is to release an attribute (an atomic information) about a user only if at least $z-1$ other users have presented the same attribute in a past time window. $z$-anonymity is weaker than $k$-anonymity since it does not work on the combinations of attributes, but treats them individually. In this paper, we present a probabilistic framework to map the $z$-anonymity into the $k$-anonymity property.
Our results show that a proper choice of the $z$-anonymity parameters allows the data curator to likely obtain a $k$-anonymized dataset, with a  precisely measurable probability. 
We also evaluate a real use case, in which we consider the website visits of a population of users and show that $z$-anonymity can work in practice for obtaining the $k$-anonymity too.

\end{abstract}


\section{Introduction}
\label{sec:introduction}

Big data have opened new opportunities to collect, store, process and, most of all, monetize data. This has created tension with privacy, especially when it comes to information about individuals. We live in the data era, where a big part of our life is readily available in digital format, from our online activity to our location history, from what we buy to how we spend out free time~\cite{InternetData}. Recently, legislators have introduced privacy laws to regulate the data collection and market, with notable examples of the General Data Protection Regulation (GDPR) in EU, or the California Consumer Privacy Act (CCPA) in the US. 

The classical approach to publish personal information is anonymization, i.e., generalizing or removing data of the most sensitive fields. Thanks to this, Privacy-Preserving Data Publishing (PPDP) has gained attention in the last decade~\cite{fung2010privacy}.  %Born with the goal of making research datasets publicly accessible, 
It is now even more popular (and critical) with the birth of data markets where data buyers can have access to large collections of data about individuals. 
Removing the user's \textit{identifiers} (name, social security number, phone number, etc.) is not sufficient to make a dataset anonymous. Indeed, an attacker can link a user's apparently harmless attributes (such as gender, zip code, date of birth, etc.) called \textit{quasi-identifiers} (QIs) to a (possibly even public) background knowledge. In this way, the attacker can re-identify the person and gain access to further sensible information from the dataset (disease, income, etc.) called \textit{sensitive attributes} (SAs)~\cite{sweeney1997guaranteeing}. Famous is the de-anonymization of Netflix public dataset~\cite{Narayanan_Netflix_2008} based on the study of QIs.

Researchers proposed several properties that anonymized data should respect to avoid re-identification, the most popular of which is the $k$-anonymity~\cite{samarati_kanon_1998}, or \kanon for short. Despite its limits, it remains the golden standard for anonymization. \kanon imposes that the information of each person contained in the release cannot be distinguished from at least $k-1$ individuals whose information also appears in the release. \kanon is conceived for tabular and static data. In other words, the dataset must be completely available at anonymization-time. Extensions to a streaming scenario have been proposed, where continuously incoming records are processed, typically using sliding windows~\cite{cao_castle_2011}. In this case, the new records are temporarily stored, processed and released after an unavoidable delay. However, for specific applications it is fundamental to avoid any processing delay. For example, for network traffic, where it is unfeasible to store packets for a long time, or location history, if a real-time (but anonymous) stream shall be used for, e.g., mobility optimization. 

This paper proposes a novel anonymization property called $z$-anonymity, or \Anon for short. It is designed to work with data streams and can be achieved with zero-delay (hence the choice of the letter $z$ instead of $k$). We assume to observe a raw stream of data, in which users' new attributes are published in real-time as they are generated. For instance, a new transaction in their credit card, a new position of their car, or a new website they visit. These attributes are QIs, and, when accumulated over time, may allow users' re-identification.

\Anon builds on the same idea of $k$-anon. When a new attribute arrives, it is released only if at least $z-1$ individuals have presented the same attribute in the past window $\Delta t$. Otherwise, it is blurred. \Anon is weaker than $k$-anonymity since it cannot guarantee that at least $k-1$ users present the same \emph{combinations} of QIs (i.e., the aggregated record). Implementing \Anon in real-time at high speed requires ingenuity, especially considering the large number of attributes the system deals with - i.e., the high-dimensional data problem, which is one of the problems hampering \kanon too \cite{aggarwal2005k}. In this paper, we show that \Anon can be obtained both with zero-delay and in an efficient way when employing a scalable implementation and appropriate data structures. Lastly, we present a probabilistic framework to map $z$- into $k$-anon properties. We find out that \Anon can provide \kanon with desired probability, for appropriate values of $z$.

There are various examples of application of \Anon. For instance, we originally proposed it for internet traffic analysis, where high-speed passive monitors process packets that contain QIs (e.g., hostnames of visited websites) in real time~\cite{favale2020aMon}. Similarly, the user browsing history, the credit card history, and  the location history offer rich information that companies want to access as quickly as possible, i.e., datum after datum, without waiting for records to be aggregated. For instance, recent credit card transactions can be useful for fraud detection or shopping recommendations; the browsing history for personalized advertisements or market intelligence; the location history to promptly optimize the mobility, or study patterns of real-time traffic.
%\Anon suits in general any real-time applications that build on big data platforms such as Kafka Stream or Spark Streaming.

In the remainder of the paper, after presenting the related work (Section~\ref{sec:related}), we formalize the \Anon property and present an approach to implement it efficiently and in real-time (Section~\ref{sec:metho}). We then propose a probabilistic model to derive $k$-anon properties from $z$-anonymized streams (Section~\ref{sec:kanon}), and study the effect of the different parameters (Section \ref{sec:params}). We then apply the model to the browsing history use case (Section~\ref{sec:use_case}). Finally, we discuss the limitations of our approach and future work (Section~\ref{sec:limitations}) and draw the conclusions (Section~\ref{sec:conclusion}).
\section{Related work}
\label{sec:related}

The problem of providing anonymization guarantees to dynamic datasets arose together with the increasing attention towards PPDP. Several approaches have been proposed during the years, that we can roughly group in microaggregation, input and output perturbation, generalization and suppression, clustering-based and tree-based techniques.

Microaggregation techniques (\cite{khavkin_preserving_2018, domingo-ferrer_steered_2019}) group the data and release an aggregated version of them, so that the user's sensitive attributes are not released as is. Input perturbation (\cite{domingo-ferrer_steered_2019, chamikara_efficient_2019}) methods aim at adding noise to the incoming data, while output perturbation techniques (\cite{kim_framework_2014,abdelhameed_restricted_2019}) generally modify the output so that it is not possible to link a user by means of a sensitive attribute with high confidence.

Other methods directly emerged from the \kanon concept where a user is indistinguishable from at least other $k-1$ users in the release. Authors of~\cite{wang_two_2018} propose two algorithms using suppression and generalization to avoid a correlation analysis from items of a transaction. Achieving \kanon is not trivial with high-dimensional data where the number of possible combinations of attributes explodes. 
Popular approaches are based on trees (\cite{li_anonymizing_2008,zhou_continuous_2009,zhang_kidsk-anonymization_2010}) or clustering (\cite{su_k_2018,sakpere_adaptive_2015,cao_castle_2011,otgonbayar_k-varp_2018,otgonbayar_toward_2016}). The rationale is the same: firstly load the incoming records in a structure (either a tree or a cluster) and secondly release those tuples when \kanon is achieved, while maintaining a trade-off with information loss. 
%\kanon has been extended with the \emph{l-diversity}~\cite{machanavajjhala2007diversity} and \emph{t-closeness}~\cite{li2007t}, sharing the idea that uncommon quasi-identifiers must be hidden to prevent users' re-identification. \mt{anche qua, l-diversity e t-closeness sono mere varianti di kanon, non specifici per transactions.}

The majority of the previous methods, however, works with the concept of sliding window, i.e., the incoming data is accumulated, then processed, and finally released with a certain delay. Some efforts have been spent to reduce the delay as far as possible: authors of \cite{zhou_continuous_2009} include the delay in the concept of output quality, with a trade-off between data quality and batch size. % (a larger one allows more effective anonymization).

To the best of our knowledge, the only work that approached the problem of zero-delay anonymization is \cite{kim_framework_2014}, where the authors propose an output perturbation approach. When a sensitive attribute arrives, it is published along with other $s-1$ other different sensitive attributes, 
%\lv{Explain better. Other $s-1$ different attributes, or other $s-1$ users with the same attribute (and value)? is this output perturbation?} 
so that the attacker can find it with probability not higher than $1/s$. 
%For example (with $s=3$), if in a medical streaming record a user comes up with ``cancer'' as a sensitive attribute, in the anonymized streaming the user will be published with ``cancer, hepatitis, parkinson''. 
Here, differently, we propose that an attribute is published only if at least other $z-1$ users exhibit the same attribute in the past $\Delta t$. % Authors propose other mechanisms to reduce the amount of noise information the system produces.
In the following, we formalize the concept of \Anon, that we previously empirically adopted in the context of live packet stream monitoring~\cite{favale2020aMon}. We generalize our approach and present a probabilistic framework to observe to what extent the \Anon property allows to satisfy also \kanon. %\lv{I don't get what is "to map" a properties into another one. Is it defining conditions where both are satisfied?} \nj{It is more "see how well we can guarantee a property if we apply the rules of a second property} 
\section{$z$-anonymity}
\label{sec:metho}

\subsection{Requirements}% and problem definition}
\label{sec:requirements}

Our goal is to define an anonymization property for data organized in streams that can be achieved with zero delay. Concisely, we seek at defining an anonymization strategy with the following requirements:

\begin{itemize}
    \item \textbf{Data streams:} we assume that observations arrive continuously in a stream. As such, we shall anonymize them based on a limited view. We do not know the future data, and we can only keep a (limited) memory of the past. 
    %: storing the entire data is not feasible, just the most important pieces are maintained, so we cannot store and exploit the full history, even for a short $\Delta t$ in the past.
    \item \textbf{Zero delay:} it shall be possible to achieve the anonymization property without any delay for publishing the anonymized stream. In other words, we want to make an atomic decision. %, based only on a limited view of past observations. 
    All approaches based on the processing of batches of observations are not applicable, as they need to store and process the entire batch before the release. %We want that our approach can work at very high speed -- e.g., network traffic received at tens of Gb/s.
    %\item \textbf{High dimensional data:} users are associated with a potentially unlimited set of attributes, whose size might be significantly larger than the number of users itself. In these cases, the classical approaches based on the \emph{k}-anonymity property fall short. \lv{Number of users is not important? What if it is too unlimited? See my comment at the end of Section III}
    \item \textbf{Efficient algorithm for high dimensional data:} the anonymization property shall be achieved with an efficient algorithm, allowing deployment on high speed and a large volume of data with off-the-shelf computing capabilities. It is important to carefully build an algorithm working with efficient data structures too, in order to obtain the necessary information as quickly as possible. %These considerations arise from the fact that memory accesses are expensive and can heavily impact system performance. 
    Moreover, users might expose a large set of attributes, whose number is not known \emph{a priori}. %In these cases, the classical approaches based on the \emph{k}-anonymity property fall short. \lv{Add more details}
\end{itemize}

\subsection{The $z$-anonymity approach}
\label{sec:anon}

We work on a data stream, in which we continuously receive observations that associate users with a value of an attribute. We define an observation as $(t,u,a)$, which indicates that, at time $t$, the user $u$ exposes an attribute-value pair $a$.\footnote{Here we will use attribute and attribute-value pair interchangeably.} For example, if \emph{Sex} is the attribute, and \emph{Female} is the value assumed by the attribute of user $u$ at time $t$, then $a$ is the pair \emph{(Sex, Female)}. Attributes can be related to whatever field: a visit to a web page, a GPS location, a purchase, etc. We consider attributes $a$ as  \emph{quasi-identifiers}, while \emph{sensitive-attributes} are not present. %i.e., an attacker may leverage them to re-identify the user in the anonymized release of the data. 
We want to keep private those values of attributes associated with a small group of users. % and ease the re-identification. 
We define the property of $z$-private attribute-value as follows:

\theoremstyle{definition}
\newtheorem{definition}{Definition}
\begin{definition}
An attribute-value pair $a$ is $z$-private at time $t$ if it is associated with less than $z$ users in the past $\Delta t$ time interval.

\end{definition}

Notice that the same attribute $a$ can be both $z$-private and not $z$-private at different time $t$. 

If the anonymized dataset hides all $z$-private attribute-value pairs, it achieves \Anon.
 
\begin{definition}
A stream of observations is $z$-anonymized if all $z$-private attribute-value pairs are obfuscated, given $z$ and $\Delta t$.
\end{definition}

In other words, the attributes that are associated with less than $z$ users in the past $\Delta t$ shall be obfuscated, i.e., removed or replaced with an empty identifier. The goal is to prevent rare values of attributes to be published, thus reducing the possibilities of an attacker to re-identify a user through unusual attributes. 

\begin{figure}
    \centering
    \includegraphics[width=\columnwidth]{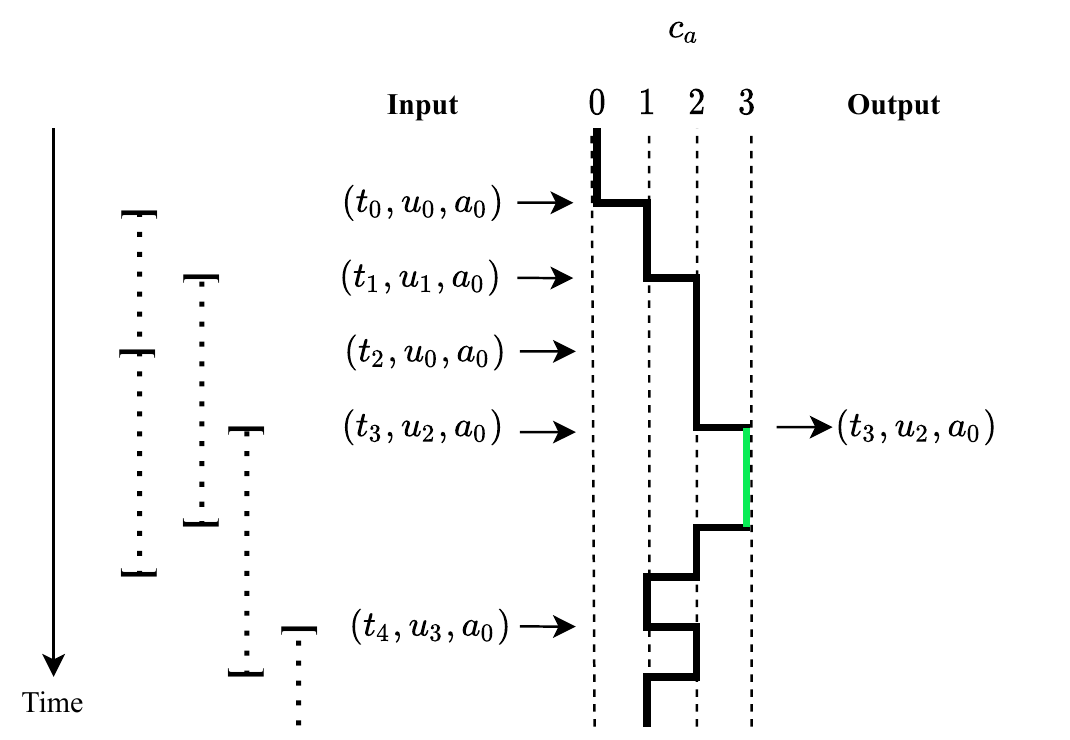}
    \caption{A graphical example of \Anon concept with $z=3$: a tuple is released only if other $z - 1=2$ different users have exposed the same attribute-value pair in the previous $\Delta t$. }
    \label{fig:alpha_diagram}
\end{figure}

We exemplify a data stream and the \Anon mechanism in Figure \ref{fig:alpha_diagram}. Assume $z=3$. At time $t_0$ user $u_0$ is the first to expose the attribute-value $a_0$.  The attribute $a_0$ is $z$-private at time $t_0$, hence it shall be obfuscated. Still, the information that $u_0$ exposed the attribute $a_0$ is kept in memory for a time equal to $\Delta t$. At time $t_1$, user $u_1$ also exposes $a_0$. Since the number of observations in $\Delta t$ is still smaller than 3, the observation is not released. At time $t_2$ user $u_0$ re-expose again $a_0$, extending the lifetime of the observation, but not changing the number of unique users having exposed $a_0$. At time $t_3$, user $u_2$ exposes $a_0$, making the total users in the past $\Delta t$ equal to 3. Thus the attribute-value pair $a_0$ is not $z$-private at time $t_3$ and the observation $(t_3,u_2,a_0)$ can be be released. At time $t_1 + \Delta t$ the attribute $a_0$ related to user $u_1$ expires, hence the total user count decreases back to 2. The same happens when $u_0$ observation expires (at $t_2 + \Delta t$), so that when $u_3$ exposes $a_0$ at $t_4$ the observation cannot be released.

In a nutshell, in a stream of incoming data, an observation is released if and only if at least $z-1$ other users had an observation for the same attribute-value pair in the past $\Delta t$ time interval. 

$z$ and $\Delta t$ are system parameters that can be tuned to regulate the trade-off between data utility and privacy. This allows \Anon to adapt to the needs of the desired use case, resulting in a flexible paradigm that can be used in many different fields. A large $z$ and a small $\Delta t$ result in the majority of attributes to be anonymized, while a small $z$ or a large $\Delta t$ allows rare values to be possibly released. $\Delta t$ regulates the memory of the system.

It is important to recall that \Anon acts in an attribute-by-attribute fashion, not considering their combinations as in the $k$-anon property. Hence, it is interesting to study which guarantees the \Anon algorithm offers in a global perspective, i.e., which assumptions it is possible to make on the overall privacy properties (e.g., it terms of $k$-anon) of the output. %Recall that while \Anon acts on a single attribute at a time, the $k$-anon is a property enforced on the whole set of quasi-identifiers.

\subsection{Implementation and complexity}
\label{sec:algo}

The \Anon property can be achieved in real-time with zero delay using a simple algorithm based on efficient data structures. We propose to generalize the approach presented in our previous work~\cite{favale2020aMon}: the attribute-value pairs $a$ are stored as a hash table $\mathcal{H}$, with linked lists to manage collisions. Each value $\mathcal{H}(a)$  in the hash table contains three elements:
\begin{itemize}
    \item metadata about $a$;
    \item a Least Recently Used list LRU$_a$ of tuples $(t,u)$;
    \item a hash table $\mathcal{V}_a$ for the users.
\end{itemize}

The idea is to minimize the time spent searching into the data structures, therefore reducing the memory accesses. By assuming that the number of attributes $a$ has the same order of magnitude of the hash structure dimension, collisions are infrequent, and consequently, the total computational cost is $O(1)$ for each incoming observation. 

The $\mathcal{H}(a)$'s metadata include the counter $c_a$ and the reference for the LRU$_a$ first and last attribute. Referring to Algorithm~\ref{alg:algo}, once an observation $(t,u,a)$ arrives, the value $a$ should be inserted in the hash table, if not already present (lines 2-6), otherwise an update should be performed (lines 7-21).
The hash value is calculated and the access to the table is done in $O(1)$.

If the user $u$ comes with attribute $a$ for the first time in the previous $\Delta t$, the user $u$ is inserted into $\mathcal{V}_a$ in $O(1)$, $c_a$ is increased by one and the tuple $(t,u)$ is inserted on top of the LRU$_a$ in $O(1)$ thanks to the aforementioned references (lines 8-11). If $u$ was instead already present in $\mathcal{V}_a$ and in LRU$_a$ with value $(t^\prime, u)$, we replace $t^\prime$ with $t$ and the tuple $(t,u)$ is moved on the top of the LRU$_a$. Again all is done in $O(1)$ (lines 12-14).

Last, to evict old entries and consequently decrease $c_a$, we traverse the LRU in reverse order: we remove each tuple $(t^\prime, u^\prime)$ where $t^\prime < t - \Delta t$, and we decrease $c_a$ accordingly (lines 17-21).
At last, if $c_a \geq z$ the observation $(t,u,a)$ is released (lines 23-24). 

\kanon has been proved~\cite{np-hard_k} an \textit{NP-Hard} problem. Differently, \Anon property can be achieved for each observation with $O(1)$ complexity with properly sized hash-tables. 
%In order to highlight the good performances of \Anon, has been proved by~\cite{np-hard_k} that , even in case of suppression of entries, so it is unfeasible for our purposes.

\begin{algorithm}
\begin{algorithmic}[1]
\caption{Pseudo code of the algorithm to implement \Anon.}
\label{alg:algo}
\small
\STATE \textbf{Input:} $(t,u,a)$
\IF {$a \notin \mathcal{H}$}
    \STATE $\mathcal{H} \gets \mathcal{H} \cup a$ \comment{new attribute: insert it for the first time}
    \STATE $\mathcal{V}_a \gets \{u\}$
    \comment{insert new user $u$}
    \STATE $LRU_a \gets (t,u)$  
    \STATE $c_a = 1$
\ELSE 
    \IF {$u \notin \mathcal{V}_a$}
        \STATE $\mathcal{V}_a \gets \mathcal{V}_a \cup \{u\}$ \comment{insert new user $u$}
        \STATE $c_a \gets c_a + 1$ \comment{add new user}
        \STATE $LRU_a \gets (t,u)$  
    \ELSE
        \STATE $(t^\prime, u) \gets (t,u)$ \comment{update timestamp of user $u$}
        \STATE move $(t,u)$ on top of $LRU_a$
    \ENDIF
\ENDIF
\STATE { } \comment{Always evict old users}
\FOR {$((t^\prime,u^\prime)$ = last($LRU_a$); 
$t^\prime < t- \Delta t$;
$(t^\prime,u^\prime)$=next$)$}    
    \STATE remove $(t^\prime, u^\prime)$ from $LRU_a$
    \STATE remove $(u^\prime)$ from $\mathcal{V}_a$
    \STATE $c_a \gets c_a - 1$
\ENDFOR
\IF {$(c_a \geq z)$} 
    \STATE {OUTPUT $(t,u,a)$ }
\ENDIF
\end{algorithmic}
\end{algorithm}

\section{Modeling \emph{z}-anonymity and \emph{k}-anonymity}
\label{sec:kanon}

We now study the relationship between the \Anon and $k$-anon properties. In particular, we quantify how a $z$-anonymized dataset could result in a \kanon release with a certain probability. Intuitively, \Anon ensures that each published value of an attribute $a$ is associated at least with $z$ users in the past time interval, while, with \kanon, any given record (i.e., the combinations of all user's attributes) appears in the published data at least $k$ times. Recall that with high-dimensional data, the set of attribute-value combinations becomes extremely high, thus making \kanon tricky to guarantee. Here we show that with a proper choice of $z$, it is possible to release data in which users are \emph{k}-anonymized.% with a controllable probability. 

We define a simple model where users generate a stream of attributes. Each attribute has a given probability of appearance that reflects its different popularity. We assume few attributes are very popular, with a long tail of infrequent attributes that may seldom appear. This often happens in real-world systems that are governed by power-law distributions~\cite{adamic2000power}.

\subsection{User and attribute popularity model}
\label{sec:model}

We consider a system in which a set of $\mathcal{U}$ users can access a catalog of $\mathcal{A}$ attributes. Let $U=|\mathcal{U}|$ and $A=|\mathcal{A}|$.

Users generate a stream of information, exposing in real-time the attribute they have just accessed. For instance, this reflects a location tracking system in which black boxes installed on a fleet of vehicles periodically exports each car location; or operating system telemetry that periodically reports which application is running; or network meters reporting which website a user is visiting. 
The system collects \emph{reports} in the form of the tuple $(t,u,a)$, i.e., at time $t$, the user $u \in \mathcal{U}$ exposes the attribute $a \in \mathcal{A}$. 
For simplicity, we assume that users are homogeneous and all reports are independent, so that the probability of getting a report, only depends on the value assumed by $a$.\footnote{We can relax this assumption, e.g., by considering classes of users. We leave this for future work.}
In particular, we assume any user $u$ exposes the attribute $a$ with a given rate $\lambda_a$, with exponential inter-arrival time. Hence, given the time interval $\Delta t$, the number of times a user exposes an attribute $a$ is modeled as a Poisson random variable $R_a$ with parameter $\lambda_a \cdot \Delta t$ ($R_{a} \sim Poisson(\lambda_a \cdot \Delta t)$).

We denote as $X_{a}$ the random variable describing whether a user exposed at least once attribute $a$ in a time interval $\Delta t$. $X_{a}$  assumes value $1$ if a user exposes $a$ in $\Delta t$,  $0$ otherwise.
We note that $X_{a} \sim Bernoulli (p^X_a)$, 
where $p^X_a$ is the probability that a user exposes attribute $a$ at least once in the past $\Delta t$. It is straightforward to compute $p^X_a$ given $\lambda_a$ and $\Delta t$ as:

\begin{equation}
    p^X_a = P[R_{a} \ge 1] = 1 - P[R_{a} = 0] = 1 - e^{-\lambda_a \cdot \Delta t}
\label{eq:lambda_to_p}
\end{equation}

\subsection{Applying \Anon}
\label{sec:applyingAnon}

We study how a stream of data modeled as above appears when released respecting \Anon. With \Anon, $z$-private attributes at time $t$ are removed. Namely, if less than other $z-1$ users are associated with $a$ in the previous $\Delta t$, the current association is blurred. We here define the event of a report $(t,u,a)$ to be published when exposed as a random variable $O_{a}$. We have that $O_{a}$ is a Bernoulli random variable with parameter $p^O_a$.

%As before, we suppose an attacker is observing the $\alpha$-anon data stream for $N \cdot \Delta t$ period of time.
%(being the current user $u$ the $z$-th one accessing $a$)
%\lv{Leggendo questo, perche non ci semplifichiamo la vita parlando solo di $\Delta t$ e $alpha$ utenti? PASTA dovrebbe valere, quindi agli arrivi di Poisson, coincidono le medie temporali. Sbaglio qualcosa? A occhio sotto in eq (4) non riesco a vedere se sia la stessa cosa}

%where $Y_a$ is a random variable describing the number of users that are associated with attribute $a$ in the past interval $\Delta t$. 
%The sum of Bernoulli random variables $Y_{a} \sim \mathcal{B}(U - 1, p_a)$ where $\mathcal{B}$ is the Binomial distribution. Indeed, $Y_{a}$ is the sum of $U-1$ independent Bernoulli random variables with success probability $p_a$.

\begin{equation}
p^O_a=P[O_{a} = 1] = P\left[\sum_{v \in \mathcal{U}\setminus u} X_{a} \ge z-1\right]
\label{eq:p_o}
\end{equation}

Given our assumption of independent and homogeneous users, we are summing $U-1$ times the same random variable~$X_a$. We remove one user since we are checking the \Anon for the report $(t,u,a)$. Hence one user is already involved by construction. Since $X_a$ is a Bernoulli with success probability $p^X_a$, its sum results in a Binomial distribution, measuring the number of occurrences in a sequence of $U-1$ independent experiments  $\sum_{v \in \mathcal{U}\setminus u} X_{a} \sim \mathcal{B}(U - 1, p^X_a)$.

Starting from Equation~\eqref{eq:p_o} and using the probability mass function of the Binomial distribution we can derive $p^O_a$ as:
\begin{equation}
\begin{split}
p^O_a= 1 - \sum ^{z - 2}_{i = 0} {U-1\choose{i}}\left(p^X_a\right)^{i}\left(1-p^X_a\right)^{U-1-i} 
\end{split}
\label{eq:p_hat}
\end{equation}

%where ${U-1\choose{i}}$ represents the binomial coefficient of parameters $U-1$ and $i$. 
%Hence we obtained the explicit probability $p^O_a$ of an exposed report $(t,u,a)$ to be released and published. 

Similar to what we did in Section~\ref{sec:model}, we denote as $Y_{a}$ the random variable describing if a user published at least once attribute $a$ in a time interval $\Delta t$. 
We note that $Y_{a} \sim Bernoulli (p^Y_a)$, 
where $p^Y_a$ is simply:

\begin{equation*}
    p^Y_a = P[X_{a} = 1] \cdot P[O_a =1] = p^X_a \cdot p^O_a
\end{equation*}

The set of the random variables describing the presence or absence for all the possible attribute-value pairs $a \in \mathcal{A}$ for a user is denoted as $\bar{Y}=\{Y_{a}\}_{a \in \mathcal{A}}$. Again this is equal for all users, being them homogeneous.

\subsection{The attacker point of view}
\label{sec:attacker}

We assume an attacker observes the $z$-anonymized output streams for all users $u \in \mathcal{U}$ for a time $N \Delta t$ with $N \in \mathbb{R}^+$ (for simplicity, in our model we considered $N \in \mathbb{N}, N \ge 1$). Hence, in our scenario, the attacker can accumulate the output for a time span possibly much larger than the parameter $\Delta t$. Similarly to $Y_{a}$, we can thus define the random variable $Y^N_{a}$, that models whether a user exposed \emph{and} published attribute $a$ at least once during the total observation period $N \Delta t$. It is clear that $Y_{a}$ and $Y^N_{a}$ are strongly related.
In fact we have $Y^N_{a} \sim Bernoulli (p^N_a)$, where the parameter $p^N_a$ can be computed as follows:

\begin{equation*}
    p^N_a = [1 - (1 - p^Y_a)^N]
\end{equation*}

This is because for a user $u$ to expose and publish an attribute $a$ in the period $N \Delta t$, (s)he has to be associated with a value 1 of $Y_a$ at least in one of the $N$ periods $\Delta t$ long. 
At the end of the period $N \Delta t$, the attacker has observed $U$ users hence obtaining $U$ realizations $\overline{y^N}$ of the random variable $\overline{Y^N}=\{Y^N_{a}\}_{a \in \mathcal{A}}$ including all the possible attributes.

The attacker will not know the random variable $\overline{Y^N}$, and will observe only realizations of it. Let us denote as $y^N_a$ a realization of the random variable $Y^N_a$ and as $\overline{y^N}=\{y^N_{a}\}_{a \in \mathcal{A}}$ a realization of the random variable $\overline{Y^N}$.

\subsection{Getting to $k$-anon}

We want to check to what extent a z-anonymized stream of a user satisfies also k-anonymity property in the whole stream of  $U$ users.  Given a specific realization $\overline{y^N}$ of a user, our goal is to derive the probability to observe at least other $k-1$ users in $\mathcal{U}$ having the same realization $\overline{y^N}$. If this happens, the system lets $k$ users release the same attributes and thus they cannot be uniquely re-identified, resulting $k$-anonymized.

Let us consider first the probability that two realizations of ${Y^N_a}$ are equal. Let us denote the two realizations, related to two users $u$ and $v$, as ${y^N_a}(u)$ and ${y^N_a} (v)$.
The probability is simply $(p^N_a)^{2} + (1-p^N_a)^2$ because either both take the values of $1$, or both take the value of $0$. Remind that the users are assumed to act independently. The probability that two users have the same realization of $\overline{Y^N}$ is then the following:

%Let us consider first the probability that a user $u$ has the same value for the attribute $a$ as user $v$. This is simply  $\hat{p}_a^{*2} + (1-\hat{p}_a^*)^2$ because either both take the values of $1$, or both take the value of $0$. Remind that the users are assumed to act independently. The probability that two users have the same attribute set is then the following:

\begin{equation*}
p^Q = P\left[\overline{y^N(u)}=\overline{y^N(v)}\right]  = \prod_{a \in \mathcal{A}} \left((p^N_a)^{2} + \left(1-p^N_a\right)^2\right)
\end{equation*}

where $\overline{y^N(u)}$ and $\overline{y^N(v)}$ are the two realizations of $\overline{Y^N}$. The parameter $p^Q$ can be seen as the parameter of a Bernoulli random variable $Q$ describing whether two realizations are equal (assuming value $1$) or not (assuming value $0$).

Finally we define the probability that a given realization $\overline{y^N(u)}$ satisfies the k-anonymity property. Hence, it means that there are at least $k-1$ other users with the same realization. We denote this probability as $p_{k-anon}$. 

\begin{equation*}
p_{k-anon}=P\left[\sum_{v \in \mathcal{U}\setminus u} Q\geq k -1\right]
%p_{k-anon}=P\left[\sum_{v \in \mathcal{U}\setminus u} \mathbbm{1}_{\{\overline{y^N(u)}\}} \overline{y^N(v)} \geq k -1\right]
\end{equation*}

%where the  indicator function $\mathbbm{1}_{\{\overline{y^N(u)}\}} \overline{y^N(v)}$ is equal to 1 if and only if the two realizations are equal, hence $\overline{y^N(u)}=\overline{y^N(v)}$.
Then $p_{k-anon}$ is the probability that at least other $k-1$ realizations are equal to the one studied. Again, as in Equation~\eqref{eq:p_o}, $\sum_{v \in \mathcal{U}\setminus u} Q$ follows a Binomial distribution of  $U-1$ experiments  with probability  $p^Q$. Then we can derive $p_{k-anon}$ as in Equation~\eqref{eq:p_hat}: 

\begin{equation*}
p_{k-anon} = 1-\sum_{i = 0}^{k-2}{U-1 \choose i}{\left(p^Q\right)}^{i}\left(1-p^Q\right)^{U-1-i}
\end{equation*}

In summary, our model describes the probability that a data stream undergoing \Anon results in dataset which respects the $k$-anon property. Although we can only provide a probabilistic guarantee that the released data will be $k$-anonymized, we can study and control this probability as a function of the parameters. 
%\begin{equation*}
%P[M(\hat{O}_u=\hat{O}_v),u\ne v|\hat{O}_u ],
%\end{equation*}

%where $M(\cdot)$ indicates the number of occurrences of the relative event.
%For ease of notation, we identify $P[M(\hat{O}_u=\hat{O}_v),u\ne v|\hat{O}_u ]$ as $P[\mathcal{M}]$.

\begin{table}
    \centering
    \footnotesize
    \caption{Terminology used to model \Anon and $k$-anon.}
    \begin{tabular}{c|l}
    \hline  
        $\mathcal{U}, U$ & Set and number of users \\ \hline 
        $\mathcal{A}, A$ & Set and number of attribute-value pairs \\ \hline
        $\Delta t$ & The time interval length used for evaluating \Anon \\ \hline
        $N$ &   \begin{tabular}[c]{@{}l@{}} 
                Length of the data stream, in multiples of $\Delta t$ \\
                on which we test the $k$-anonymity
                \end{tabular} \\ \hline   
        $\lambda_a$ & Exposing rate for attribute $a$ \\ \hline
        $R_a$ & \begin{tabular}[c]{@{}l@{}} 
                Random variable counting number of times a user exposes  \\         
                attribute $a$ in $\Delta t$. $R_{a} \sim Poisson(\lambda_a \cdot \Delta t)$
                \end{tabular} \\ \hline
        $X_{a}$ & \begin{tabular}[c]{@{}l@{}} 
                Random variable representing whether a user exposes\\ 
                attribute $a$ in $\Delta t$. $X_{a} \sim Bernoulli (p^X_a)$ 
                \end{tabular} \\ \hline
        $O_{a}$ & \begin{tabular}[c]{@{}l@{}} 
                Random variable representing whether a report $(t,u,a)$\\ 
                is published when exposed. $O_{a} \sim Bernoulli (p^O_a)$
                \end{tabular} \\ \hline    
        $Y_{a}$ & \begin{tabular}[c]{@{}l@{}} 
                Random variable representing whether a user published at  \\ 
                least once attribute $a$ in $\Delta t$. $Y_{a} \sim Bernoulli (p^Y_a)$
                \end{tabular} \\ \hline
        $Y^N_{a}$ & \begin{tabular}[c]{@{}l@{}} 
                Random variable representing whether a user published at \\ 
                least once attribute $a$ in $N  \Delta t$. $Y^N_{a} \sim Bernoulli (p^N_a)$
                \end{tabular} \\ \hline
        %\rule{0pt}{3ex} $\bar{X}$ &  
        %        Set of random variables  $ \{X_{a}\}_{a \in \mathcal{A}}$  \\ 
        %         \hline   
        \rule{0pt}{3ex} $\overline{Y^N}$ &  
                Set of random variables  $ \{Y^N_{a}\}_{a \in \mathcal{A}}$  \\ 
                 \hline
        $Q$ & \begin{tabular}[c]{@{}l@{}} 
                Random variable representing whether two realizations of \\ 
                $\overline{Y^N}$ are equal. $Q \sim Bernoulli (p^Q)$
                \end{tabular} \\ \hline    
        $p_{k-anon}$ &  \begin{tabular}[c]{@{}l@{}}
                \rule{0pt}{3ex}Probability that a realization of $\overline{Y^N}$ satisfies \\
                k-anonymity property
                \end{tabular} \\ \hline      
    \end{tabular}
    \label{tab:terminology}
\end{table}

%%%%%%%%%%%%%%%%%%%%%%%%%%%%%%%%%%%%%%%%%%%%%%%%%%%%%%%%%%%%%%%
\section{Comparing \emph{z}-anonymity and \emph{k}-anonymity}
\label{sec:params}
%%%%%%%%%%%%%%%%%%%%%%%%%%%%%%%%%%%%%%%%%%%%%%%%%%%%%%%%%%%%%%%

In the following, we show the impact of the system parameters on the \kanon and \Anon properties. 
In our model, we assume a small set of popular attributes and a large tail of infrequent ones. This allows us to catch the nature of systems where users are more likely to expose top-ranked attributes, but there exist a large catalog. %As mentioned before, this is the case of many use cases. 
As such, we choose that the $\lambda_a$ for all attributes follow a power law in function of their rank. Let us suppose attributes are sorted by rank, and the most popular attribute is $a_1$ and the least popular $a_A$. We impose $\lambda_{a_1} = 0.05$ and set the remaining $\lambda_a$ as the power-law function $\lambda_{a_r} = \nicefrac{0.05}{r}$, where $r$ is the rank of attribute $a_r$. The $p^X_a$ value is evaluated as described in Equation \eqref{eq:lambda_to_p} - for the sake of simplicity, we consider $\Delta t = 1$ unit of time. Notice that the different attributes are independent and 
%the random variable $X_a$ only depends on $p^X_a$ which are independent, and, as such,
$p^X_{a_r}$ is not a distribution probability mass function, hence it does not have to sum to $1$.

We have defined a model that describes the probability that in the released data, satisfying \Anon, a user has at least $k-1$ other users with the same set of associated attributes. Formally speaking, $p_{k-anon}=\mathcal{F}(U,A,\lambda,N,z,k) \rightarrow [0,1]$. As such, $\mathcal{F}$ gives the probability a generic user is $k$-anonymized in the released data. 
Each of the above parameters has an impact on the output probability $p_{k-anon}$. Here, we study the impact of different combinations of parameters.
%For this, we run the model with different combinations of parameters to study the impact of each one to the final $p_{k-anon}$. 
Where not otherwise noted, the default parameters listed in Table \ref{tab:default_params} are used.

\begin{table}
    \centering
    \caption{The default values used for the model.}
    \normalsize
    \begin{tabular}{c|r}
        \hline
        \multicolumn{1}{c|}{\textbf{Variable}} & \multicolumn{1}{c}{\textbf{Default Value}} \\ \hline
        $U$ &  50\,000\\ \hline
        $A$ &  5\,000\\ \hline
        $\lambda_{a_r}$ & 0.05\,/\,r \\ \hline
        $N$ &  24\\ \hline
        $z$ & 20 \\ \hline
        $k$ & 2 \\ \hline
    \end{tabular}
    \label{tab:default_params}
\end{table}

\begin{figure}[t]
    \centering
    \includegraphics[width = 0.9\columnwidth]{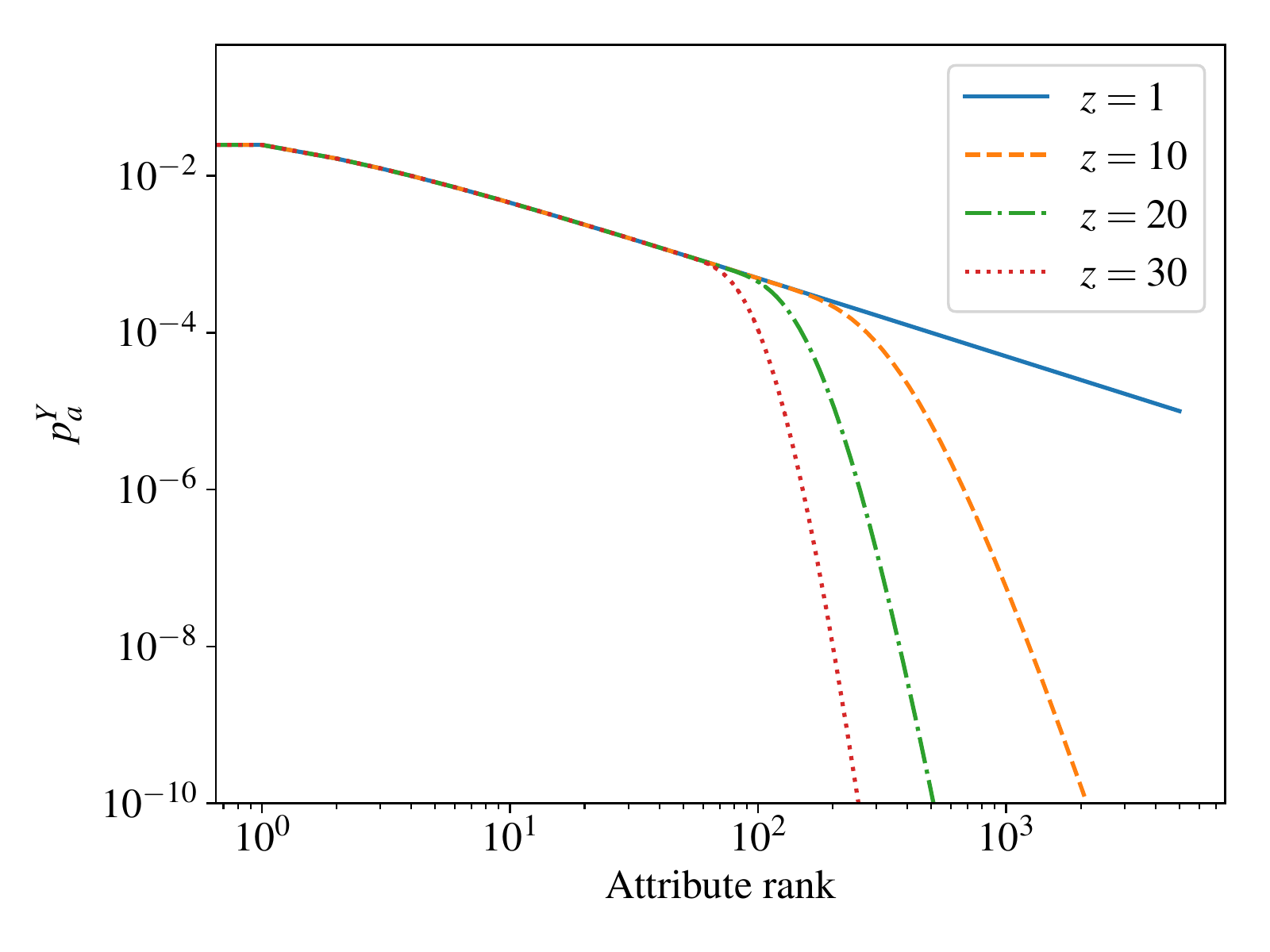}
    \caption{The probability $p^Y_a$ for a user to publish attribute $a$ in $\Delta t$, according to its rank.}
    \label{fig:pstar}
\end{figure}

\subsection{The impact of the attribute rank}
We first focus on the $p_a^Y$, i.e., the probability of observing at least once the attribute $a$ in a $\Delta t $, for a given user, in the released data, after z-anonymization. Figure~\ref{fig:pstar} shows the $p_a^Y$ in function of the attribute rank. Remind that the popularity of attributes follows a power law, since $\lambda_{a_r} \approx r^{-1}$. Indeed, the blue solid line in the figure shows the probability of observing an attribute in case $z=1$, i.e., no anonymization (equal to $p_a^X$). The curve appears as a straight line, representing a power law on the log-log plot. When enabling \Anon ($z>1$), we notice that the probability of observing uncommon attributes abruptly decreases with an evident knee. For example, if we observe the curve for $z=20$ (green dashed line in the figure), already the 300$^{th}$-ranked attribute is observed with a probability below $10^{-6}$, while it appears on the original stream with $10^{-3}$. A higher $z$ moves the knee of the curve closer to the top-ranked attributes. In other words, the figure shows how \Anon operates in preventing uncommon attributes from being released. Indeed, those attributes are released only when enough users are exposing them, hence very rarely. 
%thus preserving users' privacy. However, this happens by chance, potentially very rarely.

\subsection{The impact of $A$}
\begin{figure}
    \begin{subfigure}{\columnwidth}
        \centering
        \includegraphics[width=0.9\columnwidth]{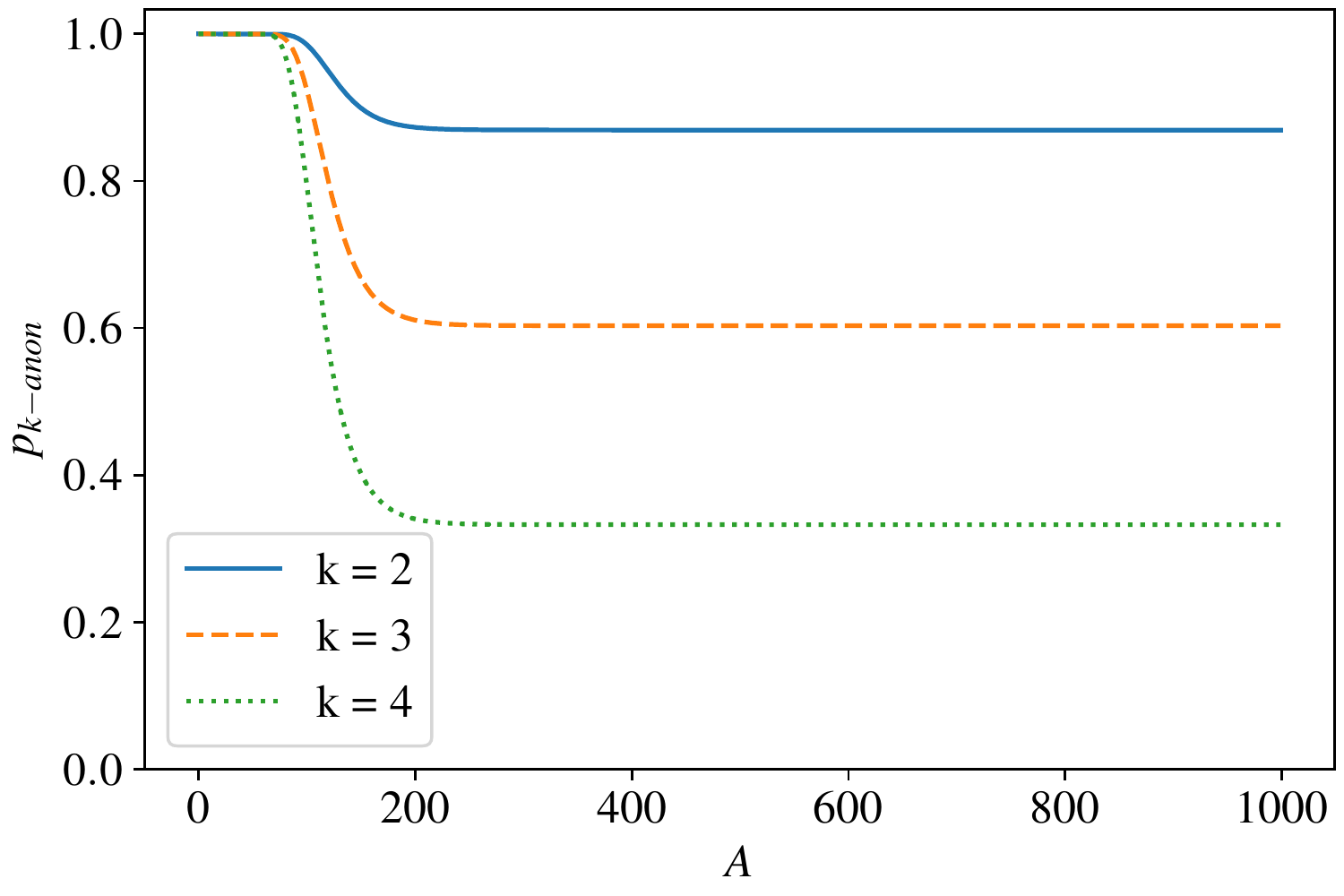}
        \caption{$p_{k-anon}$ changing $k$ ($z=20$).}
        \vspace{5mm}
        \label{fig:A_vs_k}
    \end{subfigure}
    \begin{subfigure}{\columnwidth}
        \centering
        \includegraphics[width=0.9\columnwidth]{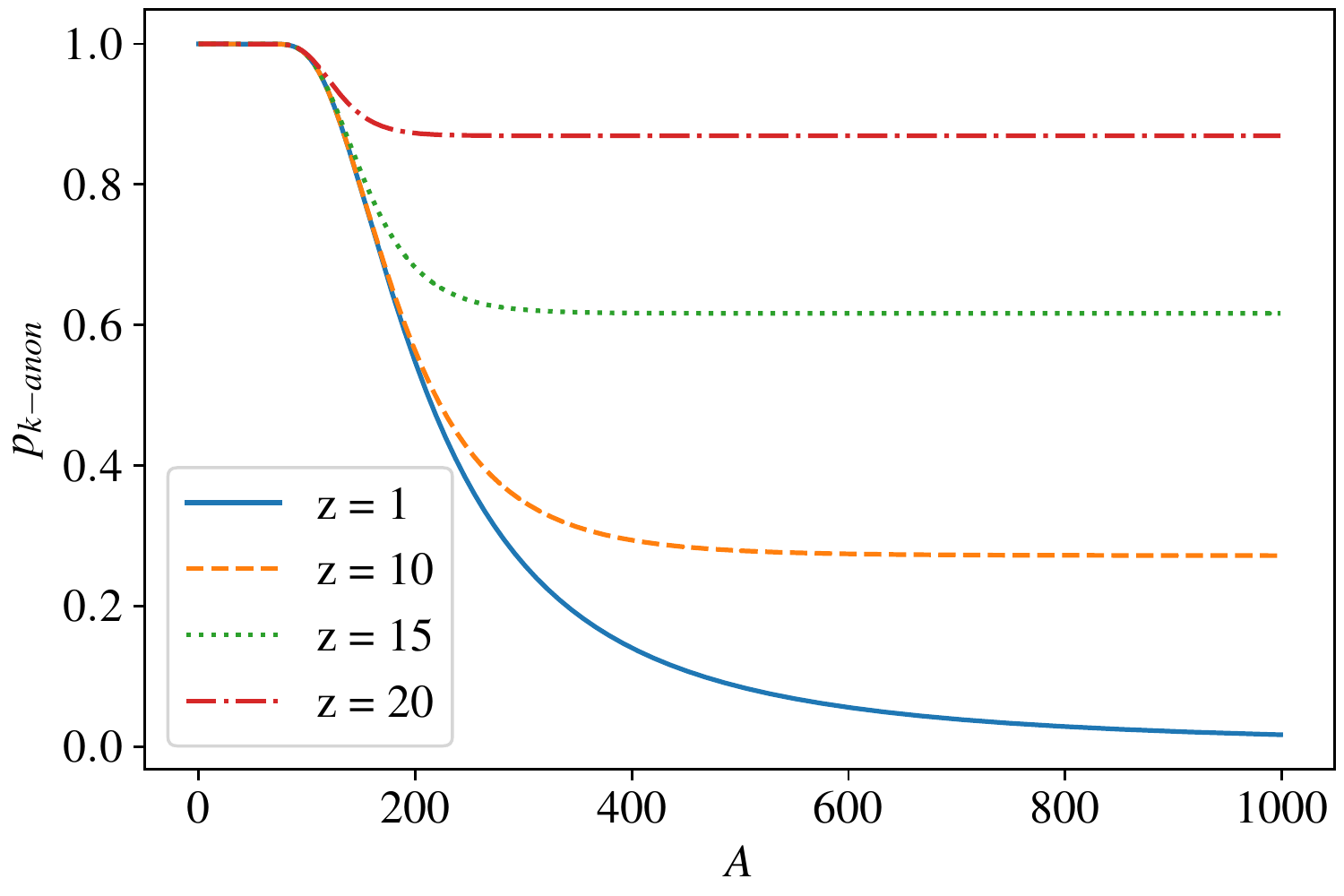}
        \caption{$p_{k-anon}$ changing z ($k=2$).}
        \label{fig:A_vs_z}
    \end{subfigure}
    \caption{The impact of $A$ on $p_{k-anon}$, considering both different $k$ and $z$ values.}
    \label{fig:A}
\end{figure}

In Figure \ref{fig:A}, we study the impact of the size of the catalog of attributes $\mathcal{A}$. In Figure~\ref{fig:A_vs_k} we show in a \Anon dataset how the probability $p_{k-anon}$ of a user being $k$-anonymized varies with $\mathcal{A}$. To this end, we perform different simulations with increasing numbers of attributes $A$. 
We consider a system where only the top $A$ ranked attributes exist. Intuitively, with a large number of attributes, it is harder to find users with the same output attribute set $\overline{y^N}$. However, our assumption of a long tail of infrequent attributes plays with us. indeed, the probability of observing them rapidly goes to to $0$ (see Figure~\ref{fig:pstar}), and, as such, these attributes rarely appear in the users' released sets. Figure~\ref{fig:A_vs_k} shows this behavior with $k = 2, 3, 4$, while keeping constant values of $z$ and $U$. With a very small catalog of top-$100$ or less attributes, users are $k$-anonymized with reasonable certainty, being very likely to observe multiple users with the same set $\overline{y^N}$. When $A$ increases, we start releasing less-popular attributes. The number of possible attribute combinations thus explodes exponentially\footnote{The attribute combinations  increase as $2^A$.}, and \Anon starts showing its effects. Focusing, for example on the orange dashed curve for $k=2$, when $A$ exceeds $100$, the probability of finding $1$ or more identical users to a given one suddenly decreases. However, it settles to approximately $0.9$ with $A>100$, clearly showing the effect of \Anon. The infrequent attributes are not released, and, as such, this limits the explosion of the possible combinations. Further enlarging $A$ does not affect $p_{k-anon}$, as the attributes in the tail are anyway not published. Increasing the value of $k$ results in lower probability of satisfying $k$-anon property. % being different values at which $p_{k-anon}$ flattens. %With $k=3 (4)$, $p_{k-anon}$, stabilizes at $0.5 (0.35)$.

For comparison, in Figure \ref{fig:A_vs_z} we report the effect of finding at least an identical user to a given one with different values of parameter $z$ of \Anon. Similarly to the other cases, $p_{k-anon}$ starts at $1$, when few attributes are present, and the number of their possible combinations is low. When $A$ increases, less frequent attributes start to appear. The possible combinations of attributes explode exponentially. With $z = 1$, i.e., no \Anon in place, the probability of finding identical users rapidly goes to 0. Enabling \Anon, we prevent rare attributes to be released, thus reducing the possible combinations. The higher $z$, the higher the $p_{k-anon}$.

%Note that in these experiments $U$ and $z$ are fixed. 
%In summary, \Anon prevents the long tail of attributes from making the realization of $k$-anonymity very unlikely.
In summary, \Anon allows $k$-anonymity to be satisfied with a non-zero probability, even with a  long tail of attributes. % very unlikely prevents the long tail of attributes from making the realization of $k$-anonymity very unlikely.

\subsection{The impact of $z$}
\begin{figure}
    \centering
    \includegraphics[width = 0.9\columnwidth]{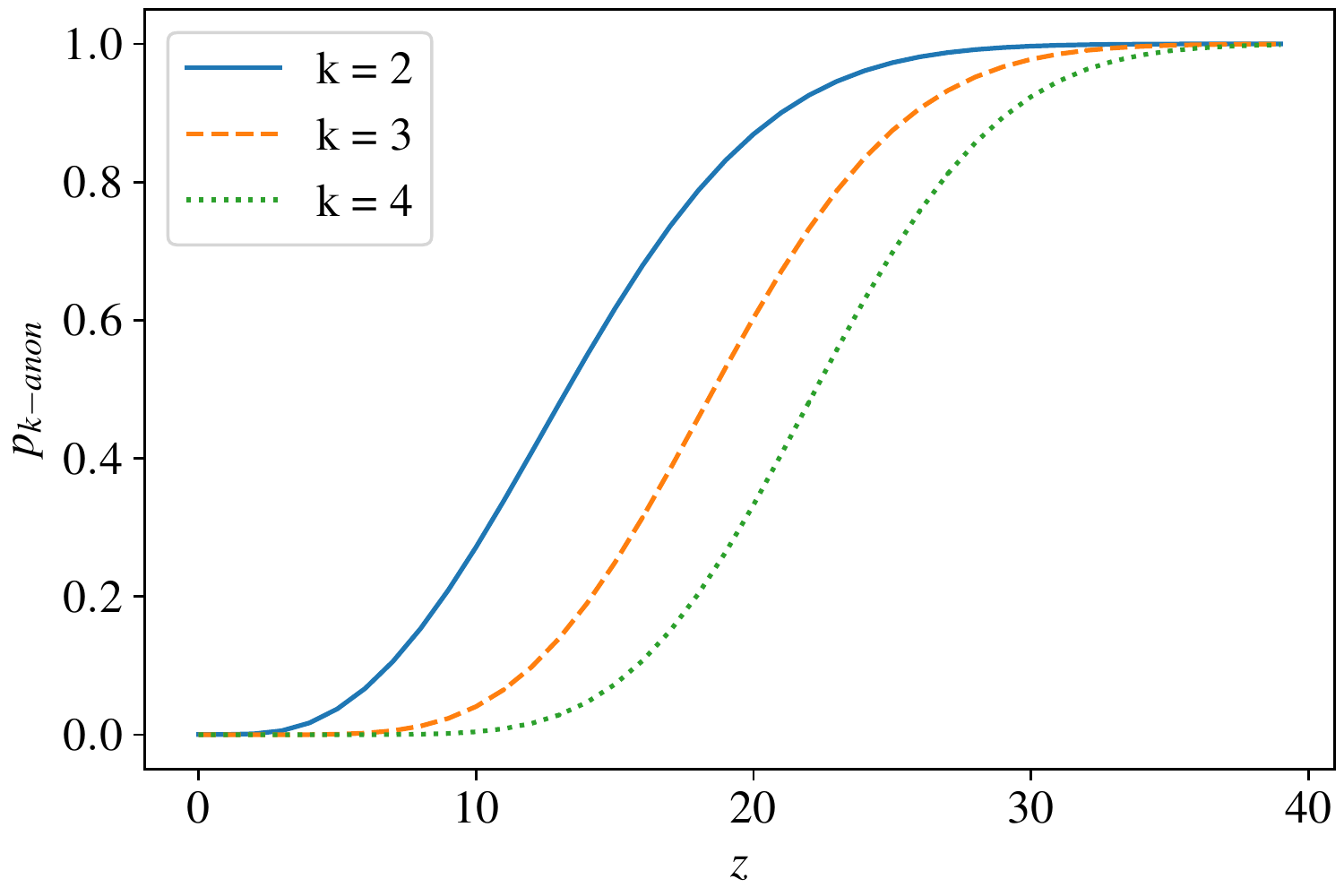}
    \caption{The impact of $z$ on $p_{k-anon}$ for different $k$ values. }
    \label{fig:zeta}
\end{figure}

We now evaluate the impact of $z$ on the $p_{k-anon}$. In Figure~\ref{fig:zeta}, we report how different values of $z$ result in different probabilities for a given user to be $k$-anonymized, i.e., there are at least $k-1$ other users with the same set of released attributes. The other parameters are fixed to the values shown in Table \ref{tab:default_params}, and different lines correspond to different values of $k$. Intuitively, the larger is $z$, the higher is $p_{k-anon}$. Focusing on $k=2$ (blue solid line),  $p_{k-anon}$ increases starting from $z=4$. With $z=20$, the probability of finding at least a user with an identical set of released attributes is already $0.8$. When $k>35$, $p_{k-anon}$ approaches $1$ for the three curves, giving the almost certainty that the whole release is $k$-anonymized (for $k=2,3,4$). %A more restrictive $k$ mandates a higher $z$ to ensure the \kanon. With $k=3 (4)$, represented by the orange (green) dashed (pointed) line, it is necessary to set approximately $z=35 (40)$ to obtain a $p_{k-anon}$ close to $1$.
In other words, it is possible to choose a proper $z$ to enforce a desired $k$ and $p_{k-anon}$ on the released data.%Please notice that, as we will show in section \ref{sec:use_case}, the $z$ is the only parameter a system administrator would be able to manage to achieve a probability of obtaining \kanon, as $U$, $A$, $\lambda$, $N$ would be imposed by the context.

\subsection{The impact of $U$}

\begin{figure}[t]
    \centering
    \includegraphics[width = 0.9\columnwidth]{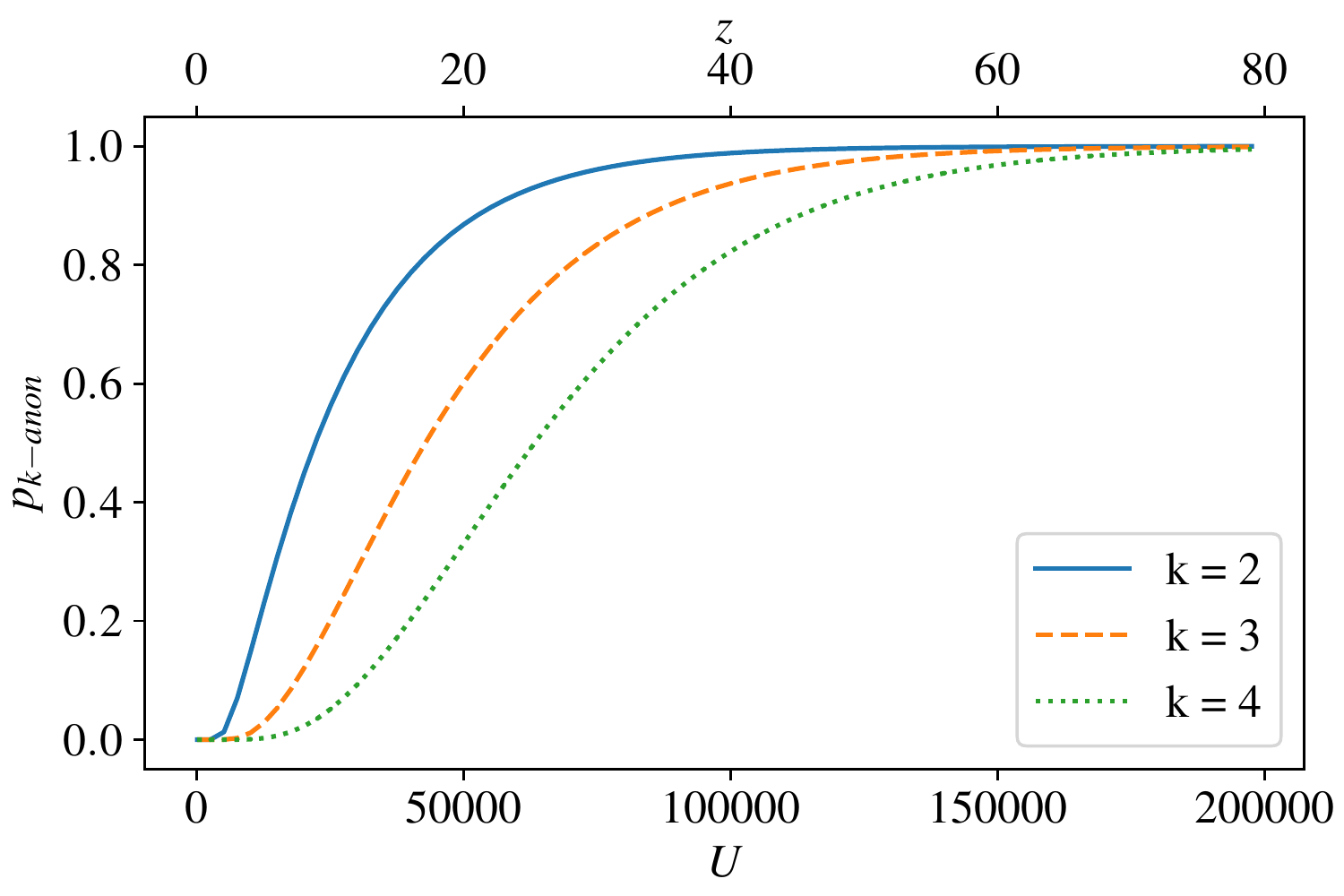}
    \caption{The impact of $U$ and $z$ on $p_{k-anon}$ for different $k$ values ($z=20$).}
    \label{fig:u}
\end{figure}

Next, we study in Figure~\ref{fig:u} how the number of users $U$ impacts the privacy of the released data. 
If we only increase the number of users $U$, not shown in the Figure, there is a higher chance that some users have even rare attributes released, breaking thus \kanon. 
%Intuitively, on the one hand, the larger is the number of users, the more likely it is to find $k$ users associated with the same set of attributes, assuming they behave independently. On the other hand, the higher the chance to observe some users whose even rare attributes are released, breaking thus \kanon. 
This would happen because a large number of users would cause even less-popular attributes to overcome the $z$ threshold, increasing the number of possible combinations, and decreasing $p_{k-anon}$. Hence, for a fair comparison, $z$ is set proportional with $U$, and we report it on the upper $x$-axis of Figure~\ref{fig:u}. Again, $A$ is fixed to $5\,000$. Focusing on $k=2$ (blue solid line), we notice how $p_{k-anon}$ grows quickly with $U$. With $U=22\,000$ (and $z=9$), the probability of a user of having another user with identical attributes is already $0.5$. $p_{k-anon}$ keeps growing, even if at a lower pace, reaching value very close to $1$ with $U=100\,000$. This result shows that a large number of users leads to better guarantees of \kanon as far as $z$ is set proportionally to $U$.

\subsection{The impact of $N$}
\begin{figure}
    \begin{subfigure}[b]{\columnwidth}
        \centering
        \includegraphics[width=0.9\columnwidth]{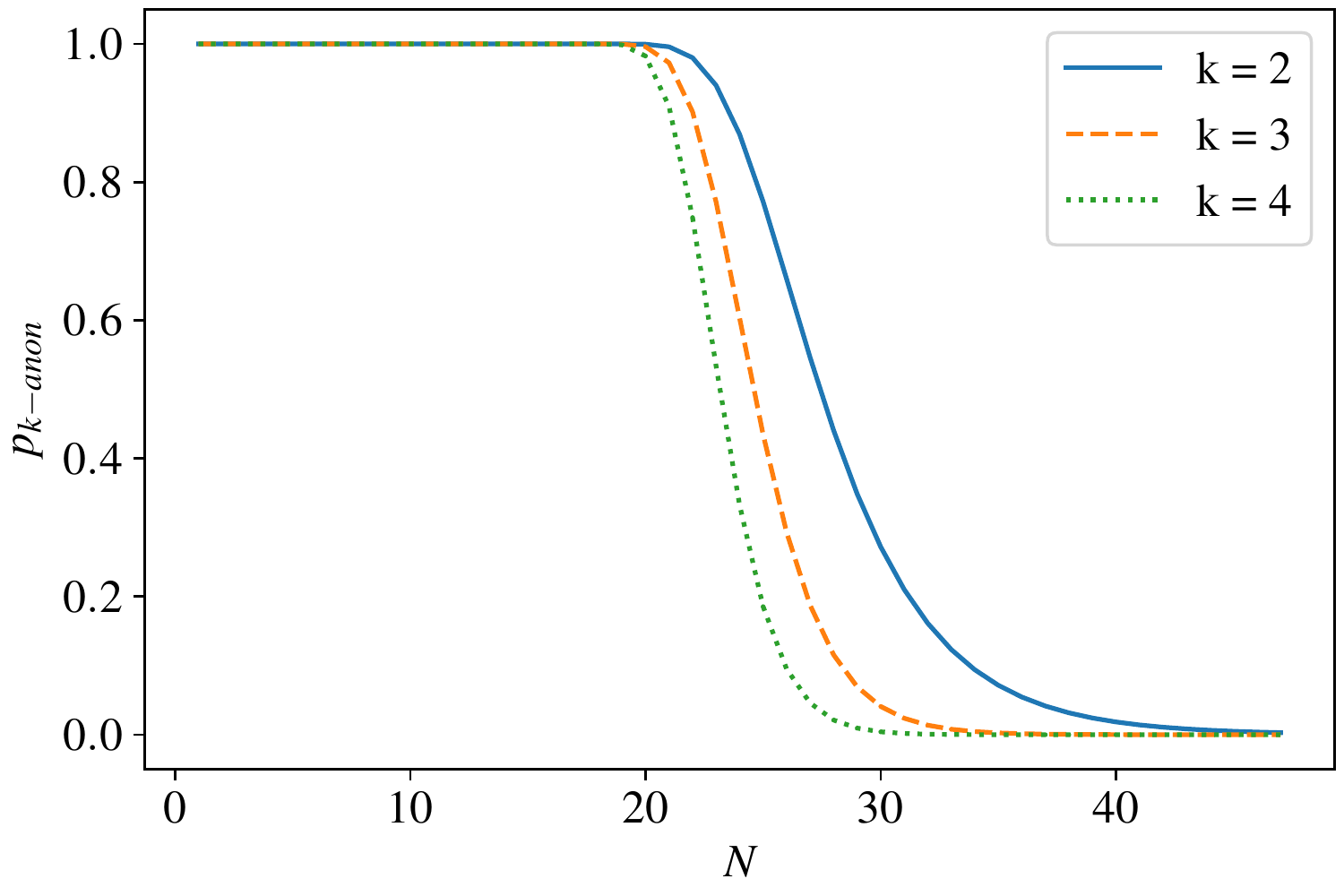}
        \caption{$p_{k-anon}$ changing k ($z=20$).}
        \vspace{5mm}
        \label{fig:N_vs_k}
    \end{subfigure}
    \begin{subfigure}[b]{\columnwidth}
        \centering
        \includegraphics[width=0.9\columnwidth]{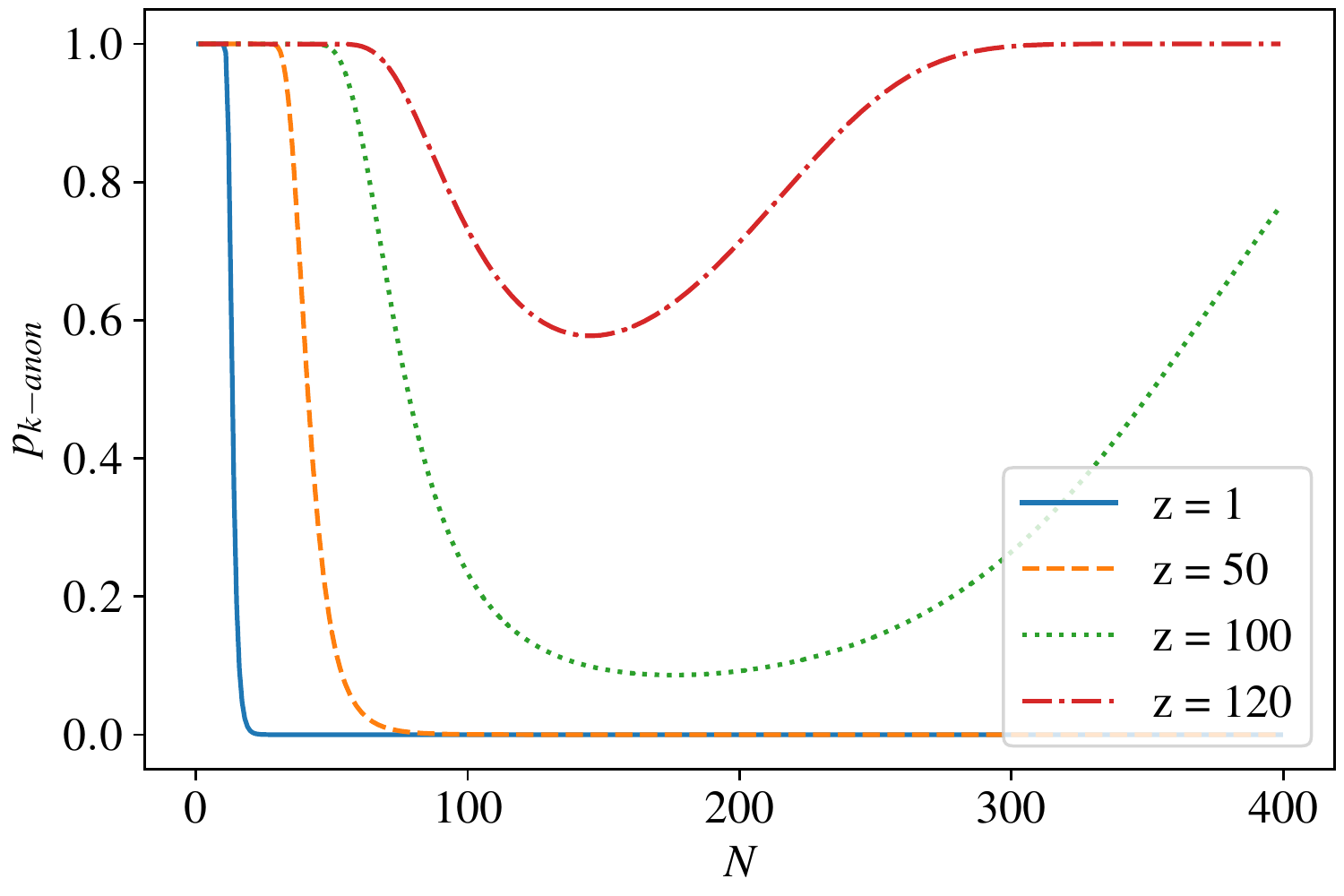}
        \caption{$p_{k-anon}$ changing $z$ ($k=2$).}
        \label{fig:N_vs_z}
    \end{subfigure}
    \caption{The impact of observation time $N$ on $p_{k-anon}$, considering both different $k$ and $z$ values.}
    \label{fig:N}
\end{figure}

Finally, Figure~\ref{fig:N} shows the impact of the observation time of the attacker ($N$), defined for simplicity in multiples of $\Delta t$. The figure quantifies how increasing $N$ affects $p_{k-anon}$.  In Figure~\ref{fig:N_vs_k} $N$ varies on the $x$-axis, while different lines represent different $k$. Intuitively, having a larger observation time makes it more difficult for users to be $k$-anonymized, since the probability that rare attributes are released increases, and, thus, the number of attribute combinations. When an attacker can access enough $z$-anonymized data, $p_{k-anon}$ drops.  Looking at the blue solid line for $k=2$, after $N=22$ periods of $\Delta t$, the probability of finding identical users starts falling, reaching $0$ with $N=45$. We observe a similar behavior with higher values of $k$ (dashed lines), for which the decrease starts earlier and it is steeper.

Figure \ref{fig:N_vs_z} shows different insights, observing the impact of the attacker obtaining data in a longer time window. Here, we fix $k=2$, and we draw different lines for different $z$, with $N$ up to $400$. With $z=1$, no \kanon can be guaranteed as soon as the attacker observes the data for $N>3$. \Anon preserves \kanon for longer time (e.g., up to $N=70$ for $z=120$). This suggests to use $z$-anon in combination with other privacy preserving approaches, e.g., user ID rotation or randomization after $N \Delta t$ time. Interestingly, with larger values of $z$, $p_{k-anon}$ grows again as the observation time increases. This happens because, sooner or later, the most popular attributes will be exposed \emph{and} published by almost every user. Hence, the observations $\overline{y^N(u)}$ will be mostly composed of $1$s, and thus most likely be equal to others. For this phenomenon to occur within a reasonable observation time, $z$ must be large enough to just consider most popular attributes, that will take less time to be exposed by almost every user.

\section{A practical use case: the visits to websites}
\label{sec:use_case}

In this section, we explore a practical use case for the \Anon: the users' navigation data. To this end, we use the data gathered on a real network to set the parameters of our model. We build on passive measurements collected by Tstat~\cite{trevisan2017traffic}, a passive meter that collects rich flow-level records, including hundreds of statistics on the monitored traffic. Essential to our analysis, Tstat builds a log entry for each TCP connection observed on the network, and, for each, it reports, among other statistics, the IP address of the client, a timestamp and the domain name of the server as indicated on the HTTP or TLS headers.\footnote{In case of HTTP transactions, the domain name is extracted from the \texttt{Host} HTTP header, while in case of the TLS from the SNI header in the Client Hello message.} We use the entries collected over one day in 2018 in a Point Of Presence of a European ISP aggregating the traffic of approximately $10\,000$ households. 
To filter those websites carrying very little information, such as content delivery networks, cloud providers or advertisement, we keep only those websites included in the top-1\,Million rank by Alexa\footnote{\url{https://www.alexa.com/topsites}} and not belonging to the aforementioned categories. For privacy reasons, we encrypted the client identifiers, i.e., the IP addresses, with the Crypto-PAn~\cite{fan2004crypto} algorithm, rotating the encryption keys every day.

We use 1 day of collected data to estimate values of the parameters. We assume $\Delta t = 1 \,hour$ and $N=24$.  We obtain $A=27\,482$ and $U=9\,670$, and we estimate directly the $p_a^X$ for each attribute (a website in this case).\footnote{We opt to extract directly the $p_a^X$ rather than $\lambda_a$ since these were directly available in the collected data.} 
Then, we setup our analysis with these obtained parameters, running our probabilistic framework and showing the results we obtain. 

%a number of unique users visiting a domain in a time interval $\Delta t$. Considering a $\delta t$ value of one hour and data from the same day as above:
%$$\lambda_{l} = \frac{\sum_{\delta = 0}^{24} \lambda_{l\delta}}{24},$$
%where $\lambda_{l\delta}$ is the domain-wise arrival rate for each specific hour, as extracted from the dataset. \mt{Rivedere questa parte.} The $\lambda_a$ values follow approximately a power law in function of the attribute rank. 

\begin{figure}[t]
    \centering
    \includegraphics[width = \columnwidth]{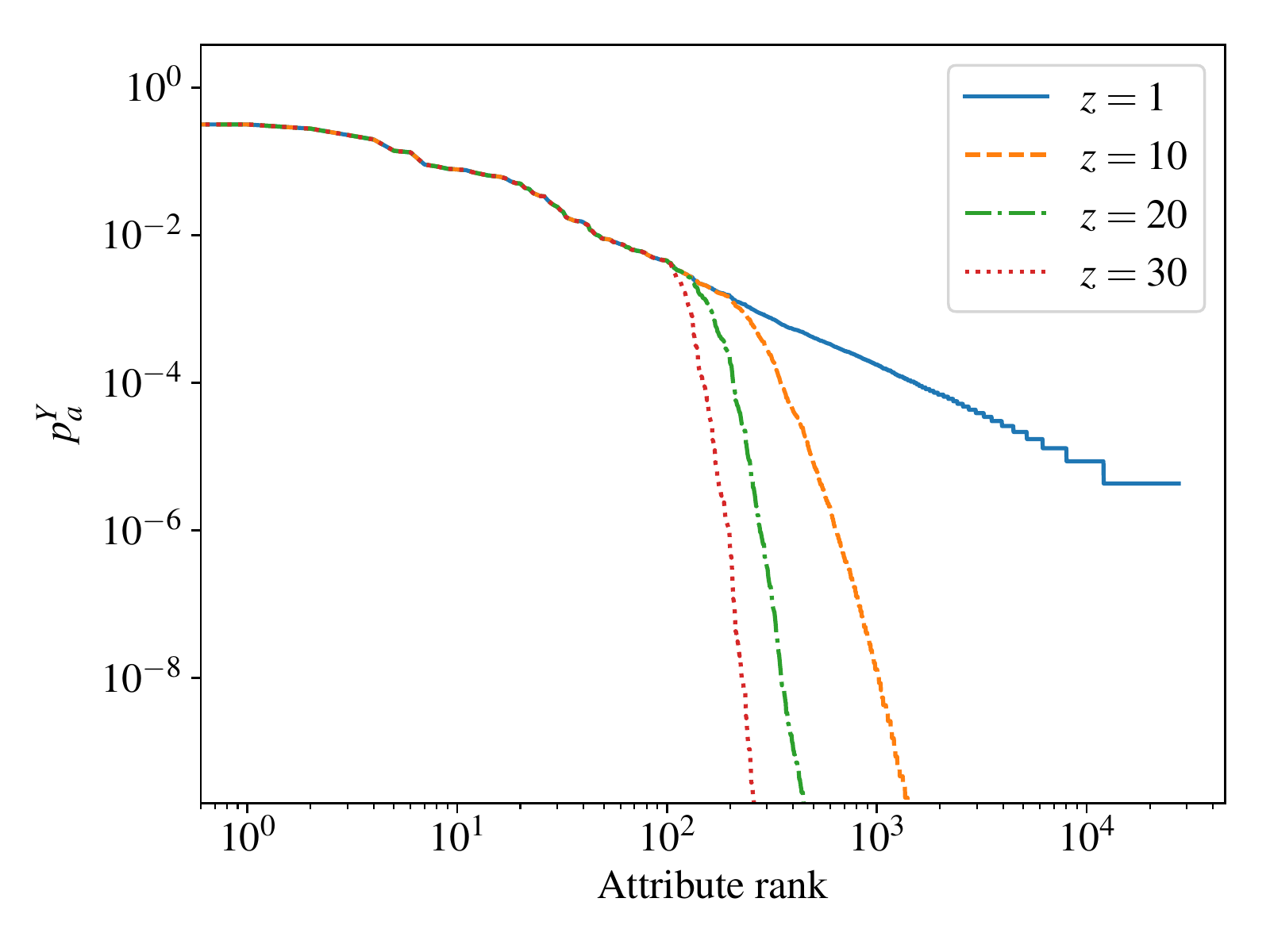}
    \caption{The probability $p^Y_a$ to publish attribute $a$ in a $\Delta t = 1 \,hour$, according to its rank, as estimated from the users' navigation data ($U = 9\,670$, $A=27\,482$).}
    \label{fig:pa_real}
\end{figure}

In Figure~\ref{fig:pa_real}, we show the probability $p_a^Y$ of observing the attribute $a$, for a given user, in the released \Anon data. The solid blue line corresponds to $z=1$, i.e., no anonymization, thus reporting the popularity of websites in the dataset. The most popular website is \emph{google.com}, which has $p^X_{google.com} = 0.34$, meaning that in 1 hour any of the users will visit this website at least once with this probability. There are some very popular websites, with the top-7 ranked having $p_a^X > 0.1$. In the tail, we find $15\,464$ websites accessed by only one user on the considered day. When running \Anon with $z>1$, these uncommon websites are not released, as they are associated with less than $z$ for most of $\Delta t$. Focusing on the orange dashed line for $z=10$, starting from the $200^{th}$-ranked website, the probability of observing it in the released data falls rapidly (notice the log scale). Higher values of $z$ (green and red dashed lines) result in earlier and steeper  decrease of $p_a^Y$. We can compare this figure with Figure~\ref{fig:pstar}, which shows the same results for the previous case. We first notice that the dashed lines (for $z>1$) move away from the solid blue line ($z=1$) in the same range $10^2 - 10^3$. Secondly, we notice that the top-ranked attributes have higher $p_a^Y$ than the previous case, with $70$ websites having $p_a^Y>10^{-2}$. This is a peculiarity of the web ecosystem, characterized by a few tens of very popular websites, including popular search engines, news portals and productivity suites, and a long tail of niche websites. In the following, we show that \Anon also works for this scenario, despite the large number of popular websites boosting the number of possible attribute combinations.

\begin{figure}[t]
    \centering
    \hspace*{0.4cm}\includegraphics[width = \columnwidth]{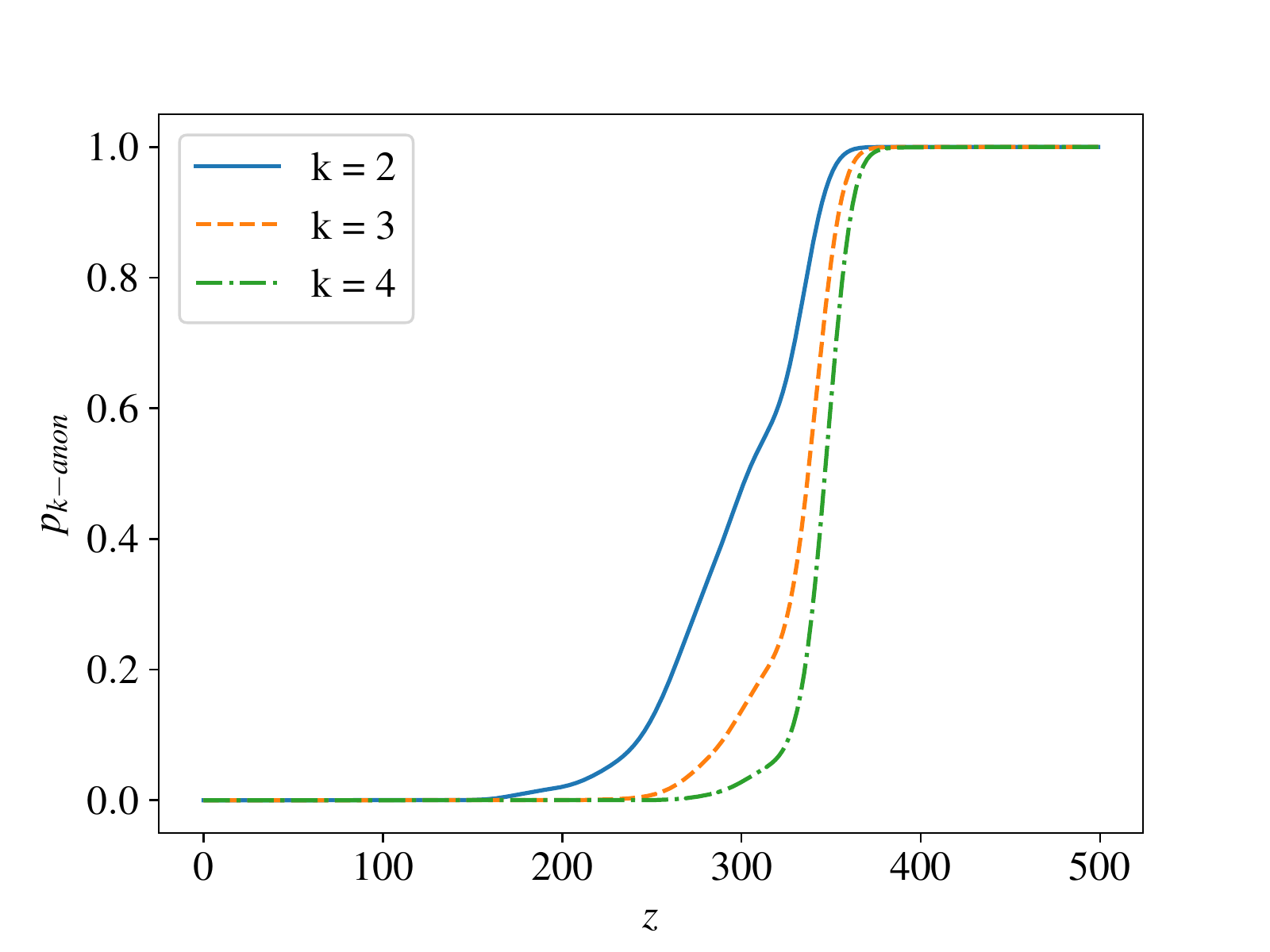}
    \caption{The relation between \Anon and \kanon in the users' navigation data ($U = 9\,670$, $A=27\,482$).}
    \label{fig:browsing_history}
\end{figure}

We now evaluate the impact of \Anon on the released data in terms of the \kanon property. Running the probabilistic framework described in Section~\ref{sec:kanon}, we can derive the probability $p_{k-anon}$ that a given user has at least $k-1$ other users with the same attribute set. We show the results in Figure~\ref{fig:browsing_history}, where we report how $p_{k-anon}$ varies with $z$, for different values of $k$. %The values $U$ and $A$ are kept fixed, as measured in the dataset. 
Focusing for example on the blue solid line (for $k=2$), we notice that $z$ must exceed $200$ for $p_{k-anon}$ to move away from $0$. $p_{k-anon}$ reaches $1$ when $z$ is $350$. When considering higher $k$ (dashed lines), larger $z$ are necessary for $p_{k-anon}$ to get close to $1$. However it is not necessary a drastic increase of $z$; for $k=4$ (green dashed line), $z=380$ is already enough. Interesting is the comparison of the website visits with the previous case study in Figure~\ref{fig:zeta}: here $z$ shall reach $380$ to obtain \kanon almost certainly, while $z=35$ is already enough for the previous case. Two reasons are behind this. Firstly, we have only $9\,760$ users for the website visits, while $U=50\,000$ in the previous case, decreasing the probability of finding users with the same set of attributes. Secondly, the probability $p_a^X$  to expose an attribute is quite  different for the two cases, with the most popular websites being visited by a large portion of users on a hourly basis. %The interesting result is that \Anon can provide reasonable guarantees of \kanon even in this case, provided it is  properly tuned. 
\Anon can provide reasonable guarantees of \kanon even in this case, provided it is properly tuned. However, this guarantees come at the cost of publishing a small number of attributes. 
This exemplifies the tension between data usefulness and privacy.

\section{Limitations and future work}
\label{sec:limitations}

With \Anon, we only prevent users' re-identification if an attacker leverages uncommon attributes, by hiding $z$-private ones. It is designed uniquely to avoid such kind of re-identification, and, so far, we do not consider other kinds of attacks, e.g.,  targeting the timing or order at which users' entries appear in the data stream. Moreover, \Anon does not consider combinations of $z$-anonymized attributes, treating them independently. Still, we provided a probabilistic framework that shows that users can be also $k$-anonymized with a controllable probability even in case an attacker knows the entire set of released attributes. With this, we provide guidelines to properly tune the system parameters to also guarantee \kanon. This allows the data curator to understand the properties of the released data and manage the trade-off between privacy and data utility. 

Future work goes in manifold directions. First, our probabilistic framework can be employed not only to assess how \Anon results into \kanon, but also to \emph{dynamically} tune $z$ to achieve a desired $k$. The probabilistic framework assumes all users behave the same. Clearly, this is a simple and strong assumption and it can be refined considering classes of users with different rates of activity as well as diverse behaviors. Moreover, in \Anon, we only considered blurring $z$-private attributes. Alternatively, we could generalize the attributes so that they pass the $z$-threshold. For example, we could generalize a website to its second level domain or its content category. Moreover, we argue that we can achieve better data utility while avoiding users' re-identification at the same time even if some $z$-private items are released. This can be obtained by introducing perturbations in the released data, e.g., by inserting noise in the data stream or modifying some of the associations between users and attributes. Such an approach melds concepts from the classical  $k$-anonymity with the ideas of differential privacy, where the addition of noise is the means to achieve users' privacy.

\section{Conclusion}
\label{sec:conclusion}

In this paper, we presented \Anon, a novel anonymization property suitable for data streams. We designed it to operate with high dimensional data, organized in transactions (atomic information about users) and with the constraint of zero-delay processing. The idea at the base of \Anon is to hide $z$-private users' attributes, i.e., those associated with less than $z-1$ other users,  which could be used by an attacker for re-identification. We show that \Anon can be achieved with an efficient algorithm if using suitable data structures. A data stream undergoing \Anon is immediately anonymized and is available with zero delay to the consumer.

\Anon is weaker than \kanon, as it operates on users' attributes independently without considering their combination. However, we provided a probabilistic framework to map $z$-anonymity into $k$-anonymity, using which the data curator can tune the trade-off between privacy and data utility. We show a practical use case, in which we evaluate \Anon using the characteristics of a real dataset of users accessing websites. We show that it is possible to tune the system parameters to obtain \kanon with a controllable probability also in this scenario.

%Our results show that a proper choice of the \Anon parameters allows the data curator to obtain a $k$-anonymized dataset with a reasonable certainty. Depending on the stream characteristics in terms of number of users and attributes, it is possible to tune \Anon to the desired privacy-data utility trade off. We show a real use case for \Anon, in which we consider the website visits of a population of users. Despite the presence of a large number of very popular websites which boost the number of possible attribute combinations, a careful choice of $z$ still allows to obtain \kanon.

%\lv{fare check references}

\bibliographystyle{ieeetr}
\bibliography{main}

\end{document}